\documentstyle[aps,preprint]{revtex}
\begin{document}
\draft
\title{\bf Stability of the black hole horizon and the Landau ghost}
\author{Jacob D. Bekenstein\footnote[1]{ Electronic
mail: bekenste@vms.huji.ac.il} and Carl Rosenzweig\footnote[2]{On leave from
Department of Physics and Astronomy, Syracuse University, Syracuse, NY}
\footnote[3]{Electronic mail: rosez@suhep.phy.syr.edu}}
\address{\it Racah Institute of Physics, Hebrew University of Jerusalem, Givat
Ram, Jerusalem 91904, Israel}
\date{Received \today}
\maketitle
\centerline{gr-qc/yymmddd}
\begin{abstract}

The stability of the black hole horizon is demanded by both cosmic censorship
and the generalized second law of thermodynamics.  We test the consistency of
these principles  by attempting to exceed the black hole extremality
condition in various process in which a U(1) charge is added to a nearly
extreme Reissner--Nordstr\"om black hole charged with a {\it different\/}
type of U(1) charge.  For an infalling spherical charged shell the attempt is
foiled by the self--Coulomb repulsion of the shell.  For an infalling
classical charge it fails because the required classical charge radius
exceeds the size of the black hole. For a quantum charge the horizon is
saved  because in order to avoid the Landau ghost, the effective coupling
constant cannot be large enough to accomplish the removal. \end{abstract}


\section{INTRODUCTION}

Notwithstanding various possible exceptions, the principle of cosmic censorhip
is a popular tenet of belief in black hole physics.  This principle rules
that  the black hole event horizon cannot be removed because that would
expose naked singularities to distant observers.  Likewise, the disappearance
of the event horizon would violate the generalized second law of
thermodynamics inasmuch as the horizon area is associated with entropy which
would thereby disappear without  any obvious way to compensate for its loss.
For these reasons processes which seem to have a chance of eliminating the
event horizon must be unphysical.  Devising candidate processes and finding
out how they fail turns out to be a source of physical insight into black
holes, and even into more mundane physics.

In this paper we inquire into the physics that defends the horizon
from attempts to transcend the extremality condition for a
Reissner--Nordstr\"om black hole.  As is well known, for such a black hole
the charge must not exceed the mass (in units with $G=c=\hbar=1$); otherwise
the Reissner--Nordstr\"om solution contains no horizon.  Attempts to violate
this condition by adding to a nearly extreme Reissner--Nordstr\"om  black
hole a particle with charge of the same sign as the hole's and with
charge--to--mass ratio larger than unity are known to be defeated by the
Coulomb repulsion.  In fact, the energy required to get the particle to
surmount the potential barrier surrounding the black hole is found to be
enough to make the mass of the hole grow more than its charge, so that the
hole becomes further removed from extremality.

But suppose there exist two types of local charge, type--$\varepsilon\in
U(1)$ and  type--$q\in U'(1)$, {\it e.g.,\/} electric and magnetic charge,
which always reside in different particles.  The black hole is assumed to
contain a total $U(1)$ charge $\epsilon$ which is close to its mass $M$, but
no $U'(1)$  charge to start with, so that it is not endowed with a $U'(1)$
gauge field.  Thus  an infalling $q$--type charge encounters no repulsive
electrostatic potential barrier and, on first sight, is not  hindered from
crossing the horizon.  Now, for two charge types the condition for the
Reissner--Nordstr\"om horizon to continue to exist after the assimilation is
\begin{equation}
 \epsilon^2+q^2 \leq M^2
\label{horizonexists}
\end{equation}
What physics prevents the added charge $q$ from violating this condition ?

We shall study two distinct {\it gedanken\/} experiments.  In the first a
$U'(1)$ charge $Q$ is let fall on the black hole as a spherical shell
concentric with the black hole.  The calculation can be carried out exactly,
and shows that, in fact,  if the shell's charge is large enough to lead to a
violation of Eq.~(\ref{horizonexists}), the shell's own self repulsion
prevents it from reaching the black hole.  This is an extension of the usual
mechanism that safeguards the horizon with one kind of charge present.

In our second {\it gedanken\/} experiment we consider an infalling
pointlike $q$--type charge of mass $\mu$.  Again it meets no repulsion from
the hole's field, but neither does self--repulsion play any visible role in
preventing its assimilation by the black hole.  In fact, we find that the
condition for transcending Eq.~(\ref{horizonexists}) is precisely that the
classical charge radius $r_c=q^2/\mu$ of the charge be bigger than the black
hole.  Thus if the particle is classical, it cannot get into the black hole,
and the attempt fails.  If the particle is an elementary quantum charge, its
size is set by the Compton length $1/\mu$. The condition for removal of the
horizon then translates into  $q^2 > 2$,  meaning that the $U(1)$ gauge
theory must be strongly coupled.

As is well known, the vacuum polarization required by QED or its analogs makes
the charge associated with a particle significantly dependent on the length
scale at which it is is looked at: at large distances most of the charge is
screened.  The condition that, on the scale of the black hole horizon, $q^2 >
2$  means that the the Landau ghost would show up at measurable scales, an
intolerable situation.  Thus if we require that the $U(1)$ gauge theory in
question be described by a consistent  effective theory,  the conditions for
removal of the black hole horizon by addition of $q$--type charge cannot be
fulfilled. The event horizon is truly stable.

\section{INFALL OF CHARGED SHELL}

The Reissner--Nordstr\"om metric \cite{MTW} must be the exterior metric of
a spherical distribution with two different $U(1)$ charges, $Q$ and
$\epsilon$ \cite{Bekenstein}   \begin{equation}
ds^2=-\left(1-{2M\over r}+{(\epsilon^2+Q^2)\over r^2}\right)dt^2+{dr^2\over
1-{2M\over r}+{(\epsilon^2+Q^2)\over r^2}}+r^2(d\theta^2+ \sin^2\theta d\phi^2)
\label{RNmetric} \end{equation}
displays an event horizon only if condition (\ref{horizonexists}) holds.
The electric potential of the $q$--type charge is \cite{MTW}
\begin{equation}
\Phi=Q/r.
\label{potential}
\end{equation}

Let us start with a black hole of mass $M$ and $\varepsilon$--type
charge $\epsilon$, but with vanishing $q$--type charge, and consider the radial
infall into it of a thick spherical shell concentric with it.  The shell is
made up of identical particles, each  bearing $q$--type charge with specific
charge $q/\mu$.  The total $q$--type charge of the shell is $Q$ and its total
rest mass  $m$.  The initial conditions are that the shell starts off  at very
large distance $r$ from the hole, and with each of its constituents having the
same given inward velocity.  Consequently, the specific energy at infinity is
a fixed quantity $E>1$ for all particles.   We neglect pressure in the shell,
{\it i.e.,\/} we assume random velocities remain negligible.  This means that
the shell has conserved energy $mE$ in the field of the black hole.

The equation of motion of a particle at the outer edge of the shell is that of
a point charge with specific charge $q/\mu$ moving in the metric
(\ref{RNmetric}) with mass $M'=M + mE$, the mass of the hole plus shell, and
in the potential $\Phi(r)$ of the shell itself.  If $\tau$ denotes the
charge's proper time, the conservation of its specific energy $E$ is written
as \cite{MTW}   \begin{equation} \left(1-{2M'\over r}+{(\epsilon^2+Q^2)\over
r^2}\right){dt\over d\tau} +{(q/\mu)Q\over  r} = {\rm const.} = E
\label{energy}
\end{equation}  This equation may be used to eliminate $dt/d\tau$ in the
normalization of the velocity \begin{equation} -\left(1-{2M'\over
r}+{(\epsilon^2+Q^2)\over r^2}\right)({dt\over d\tau})^2 + {(dr/d\tau)^2\over
1-{2M'\over r}+{(\epsilon^2+Q^2)\over r^2}}=-1 \label{normalization}
\end{equation}
The result is
\begin{equation} \left(dr\over d\tau\right)^2- {2\over
r}\left[M+mE\left(1-{Q^2\over m^2}\right)\right]+ {1\over
r^2}\left[\epsilon^2+Q^2\left(1-{Q^2\over m^2}\right)\right]=E^2-1>0
\label{firstquadrature}
\end{equation}
where we have replaced $q/\mu\rightarrow Q/m$, as well as $M'\rightarrow M+mE$.
This first quadrature for the problem has the form of an energy conservation
equation.  We refer to the terms following $(dr/d\tau)^2$ as the potential.

Two cases are of interest. (a) the shell proceeds to fall into the
black hole without any of its component shells turning back--no shell
crossing--and without transcending condition (\ref{horizonexists}). (b) The
condition (\ref{horizonexists}) is transcended.  In case (a)
Eq.~(\ref{firstquadrature}) must have a solution with $r(\tau)$ crossing the
horizon $r_H$ of the complete system:
\begin{equation}
r_H=M+mE+[(M+mE)^2-\epsilon^2-Q^2]^{1/2}
\label{rplus}
\end{equation}
If case (b) with consequent destruction of the horizon is to be possible, the
potential barrier should not be able to turn $r(\tau)$ back.  Let us consider
this second eventuality.

The black hole existed to start with, so $\epsilon\leq M$.  No horizon will
exist after assimilation of the shell if $(M+mE)^2<\epsilon^2+Q^2$. Combining
these inequalities tells us that \begin{equation}
(Q/m)^2>E^2+2EM/m
\label{over}
\end{equation}
Thus the specific charge $Q/m=q/\mu$ must be large. It also follows from
Eq.~(\ref{over}) that
\begin{equation}
M+mE(1-Q^2/m^2) < M(1-2E^2) <0
\label{second}
\end{equation}
and
\begin{equation}
\epsilon^2+Q^2(1-Q^2/m^2) < M^2(1- 4E^2) <0
\label{third}
\end{equation}
Thus the square brackets in Eq.(\ref{firstquadrature}) are
both negative if the shell is capable of removing the horizon.  This means
that the potential has a hump which could well block the shell from continuing
on its way into the black hole.

A simple calculation shows that the peak of the potential term is
\begin{equation}
V_{\rm peak}={[mE(Q^2/m^2-1)-M]^2\over Q^2(Q^2/m^2-1)-\epsilon^2}
\label{peak}
\end{equation}
and lies at
\begin{equation} r_{\rm peak}={Q^2(Q^2/m^2-1)-\epsilon^2\over
mE(Q^2/m^2-1)-M}={Q^2\over mE}+{{MQ^2\over mE}-\epsilon^2\over
mE(Q^2/m^2-1)-M} \label{rpeak} \end{equation} It now follows from inequalities
(\ref{over}) and (\ref{second}) that $r_{\rm peak}>2M$ so that the shell will
certainly reach  the potential barrier before reaching the original horizon.
Thus in order for the {\it whole\/} shell (with parameters capable of leading
to a removal of the horizon) to actually fall into the hole, it is necessary
for $E^2-1\geq V_{\rm peak}$.

Let us introduce the variables $\alpha$ and $\beta$ by
\begin{equation}
Q^2=m^2E^2+2EmM+m^2 \alpha; \qquad  \epsilon^2=(1-\beta)M^2
\label{delta}
\end{equation}
Trivially $0\leq \beta \leq 1$ while inequality (\ref{over})
guarantees that $\alpha>0$ in our case where the shell's paramaters are
appropriate for removing the horizon.  Using Eq.~(\ref{peak}) we write
\begin{equation}
V_{\rm peak}-1+E^2=
{(2 + E^2\beta - E^2 -\beta ) M^2+(E^2 m^2 -m^2 + 2 EmM)\alpha
\over  \epsilon^2 m^2  + Q^4 - m^2 Q^2 } \label{basic}
\end{equation}
A look at inequality (\ref{third}) shows that the denominator
here is positive.  In view of the ranges of $\alpha$ and $\beta$ and the fact
that $E\geq 1$ the numerator is also positive, making the whole expression
positive.  Thus by Eq.~(\ref{firstquadrature}) the outer edge of the shell
must reach a turning point before it reaches the maximum of the potential.
This means that part of the shell must be turned back.

Thus if the change in black hole parameters that would have resulted
from assimilation of the shell sufficed to remove the horizon, that whole shell
cannot reach the black hole.  The contrapositive of this is thus true: if the
shell's parameters are contrived so that all of it can reach the black hole,
it cannot remove the horizon.   Thus the classical process envisaged respects
the horizon's existence, cosmic censhorship, and the generalized second law.

\section{INFALL OF POINT CHARGE}

Now let a pointlike $q$--type charge of mass $\mu$ and charge $q$ fall
radially into a Reissner--Nordstr\"om hole of mass $M$ and $\varepsilon$--type
charge $\epsilon$ satisfying the second of Eqs.~(\ref{delta}).  If we treat
the charge as a classical test particle ($\mu\ll M$ and $q\ll \epsilon$), its
motion $\{t(\tau), r(\tau)\}$ will  again be described by Eqs.~(\ref{energy})
and (\ref{normalization}), but with $Q=0$ and $M'=M$ since the black hole
bears no $q$--type charge and the particle's influence on the background is
being neglected.  Combining the equations as in Sec.~I,  we find the first
quadrature \begin{equation}  (dr/d\tau)^2- 2M/r + \epsilon^2/r^2=E^2-1>0
\label{newquadrature}  \end{equation}
The particle will move inward until it bumps into the rising
potential ($\epsilon^2/r^2$ term).  The turning point is
\begin{equation}
r_{\rm turn}=M{\left[1+(E^2-1)(1-\beta)\right]^{1/2}-1\over E^2-1}
\label{turning}
\end{equation}
It is easy to see that $r_{\rm turn}< M/2$ for any choice of $E$
since $0\leq\beta\leq 1$.  Hence the distortion of spacetime due to the black
hole charge $\epsilon$ cannot prevent the particle with charge $q$ from
falling into it.

After the infall the black hole mass is $M+\mu E$. The condition for removal
of the horizon is  $\epsilon^2+q^2=M^2(1-\beta)+q^2>(M+\mu E)^2$.  Since
$E\geq 1$, $q^2>2M\mu+\mu^2+M^2\beta$.   Since we can make $\beta$ arbitrarily
small, and $\mu\ll M$, to remove the horizon we need at least that
\begin{equation}
 q^2>2M\mu
\label{newcondition}
\end{equation}
What physics prevents a particle with $q^2>2M\mu$ from accreting onto the black
 hole?  Let us consider some options.

As a charge is lowered towards a black hole, it polarizes the hole in such a
way that from far away the source of its field looks more spread out around
the hole than the particle \cite{CohenWald}.  Could this effective spreading
stop the ``dangerous'' particle from falling in ?  No.  One can view the
spreading as resulting from image charges induced on the black hole's surface
by the approaching charge.  Just under the charge the image charges are of
opposite sign.  Around the hole they are of the same sign.  Obviously, the
effect of the image charges should be to pull in the charge even more strongly
than gravity alone.  Thus this phenomenon cannot help to prevent assimilation
of the charge $q$ by the black hole.

The black hole might discharge  its $\varepsilon$--type charge  {\it a la
Schwinger\/} sufficiently rapidly to offset the push beyond extremality by the
added charge.  Schwinger--type charge emission would depend on the
$\varepsilon$--type electric field of the black hole, which is of order
$\epsilon/M^2\approx 1/M$ near the black hole. This field can tear the virtual
pairs in the $\varepsilon$ vacuum if the work done by it on an elementary
$\varepsilon$--type charge $e$ over its Compton length $1/m$ amounts to at
least the mass of a pair $2m$.  Thus Schwinger discharge will be exponentially
suppressed unless  $e/m>2mM$.  Now suppose that there exist in nature
$\varepsilon$--type elementary charges with $e/m\approx 1$.  We can then make
a black hole by collapsing a large number of these unmixed with other stuff.
In spherical collapse there is no energy loss to waves, so that
$\epsilon/M=e/m\approx 1$, and we can indeed form a nearly extreme black hole.
If the charges, whose Compton length is $1/m$, are to fit into the black hole
of size $M$, we must demand $mM>1$.  But then it is impossible to satisfy the
condition for Schwinger emission.  Thus one can imagine black holes that
cannot be saved from destruction by Schwinger--type discharge.

Hawking thermal emission preferentially carries charges of the same sign
as $\epsilon$; it might thus drive the black hole below extremality before the
added charge drove the hole over it.  But since the black hole is assumed near
extreme, its Hawking temperature is very small so that the emission is
unimportant.  For precisely the same reason, radiation pressure in the
``photons'' of the $U'(1)$ gauge field can be regarded as weak compared to
gravity, and is powerless to prevent infall of the charge $q$.

Our persistent failure to find a mechanism that prevents ingestion by the black
hole of a ``dangerous'' charge leads us to the conclusion that there must be
some basic physical reason why condition (\ref{newcondition}) cannot be
satisfied for a charge that {\it is\/} able to fall into the black hole.  For
a classical charge $q$ the reason is not far to seek.  We note that its
classical charge radius (analogous to the classical electron radius) is
$r_c=q^2/\mu$, and  condition (\ref{newcondition})  simply says that
$r_c>2M$.  A classical particle which does not contain a negative energy
density region somewhere in it must be larger than $r_c$ since the
electrostatic energy residing outside $r_c$ would already account for all of
the rest mass $\mu$.  Thus if the charge is capable of fitting in the black
hole ($r_c<M$), it cannot satisfy (\ref{newcondition}), and cannot be used to
remove the horizon.

However, for an elementary charge, {\it e.g.\/} an electron, $r_c$ is not the
measure of particle size.  In fact, for $U(1)$ charges found free in nature
(weak coupling constant $q^2\ll 1$), $r_c$ is far smaller than the Compton
length $1/\mu$, the true quantum measure of particle size.  Thus the charge
can fall into the black hole only if $r_c<1/\mu<M$.  But then condition
(\ref{newcondition}) cannot be satisfied, and we  recover our previous
conclusion that the horizon cannot be removed.

However, condition (\ref{newcondition})  together with the requirement that
the particle fit in the black hole, $M> 1/\mu$, means that we must consider
strongly coupled QED--type theory ($q^2>1$).  An elementary charge in such
theory has $r_c>1/\mu$,  and we cannot rule out condition (\ref{newcondition})
from the requirement that the particle can fit into the black hole,
$1/\mu<M$.  We thus look more carefully at what strongly coupled $U(1)$ gauge
theories are like.

The very notion of charge of a point particle in such a theory is dependent
on the lengthscale on which it is measured.  In a QED like theory, the
relation between the charge of a point particle at two different scales,
$\ell$ and $L$ with $L\gg \ell$, is given by  the result from renormalization
improved perturbation theory \cite{Itzik},    \begin{equation}
{1\over q(\ell)^2}={1\over q(L)^2}-{2\over 3\pi}\ln(L/\ell)
\label{scaling}
\end{equation}
The physics behind this relation is that at long scales (say macroscopic)
the charge is weaker because of vacuum polarization shielding of the charge at
small (microscopic) scales.  Evidently for $q(L)\neq 0$ there exists a
sufficiently short scale $\ell_{\rm L}$ at which  $q(\ell_{\rm
L})\rightarrow\infty$; this is the Landau ghost.  Of course, this behavior is
unacceptable.  One possible resolution \cite{QED} is that QED and similar
$U(1)$ theories are trivial, {\it i.e.,\/} $q\equiv 0$.  The Landau ghost does
not then appear. This is what happens for $\lambda\phi^4$ theories
\cite{Callaway}.   Another possibility \cite{Miransky} is that as $q(\ell)^2$
grows, the theory makes a transition to a new phase so that the Landau ghost
never shows up.  The new phase is characterized by massive four--fermion
interactions and seems to lack a long range force.  Since  $q(L)^2\neq 0$ in
the real world, and electrons interact via photon exchange, we can consider
QED  as an effective theory valid above some short scale cutoff.  Both
alternatives are consistent with all experimental facts because the Landau
ghost occurs at extremely short scales in QED (shorter than the Planck scale).

If a $U(1)$ gauge theory undergoes a phase transition at strong
coupling comparable to that implied by condition (\ref{newcondition}), we
cannot obviously talk about simple charged particles with their attendant
Coulomb interaction.  Though we are unable to work out the details of the
protection mechanism, the horizon will probably be safe.  If the theory is
trivial, it is certainly safe.  We are left with the possibility that the
$U(1)$ theory is an effective theory defined over some finite range of
scales.  Can a charge in such theory remove the horizon ?

In order for us to consider the charge as a quantum particle subject to the
effective field theory, that theory must be applicable at scales below
the particle's Compton length, {\it i.e.,\/} $\ell=1/(\xi\mu)$ with $\xi>1$.
On the other hand, the charge relevant for the motion of the charge in the
black hole's background must be defined on scales larger than $M$; hence we
need $L=\zeta M$ with $\zeta>1$.  Finally, for the effective theory to be
self--consistent, the Landau ghost must not appear, {\it i.e.\/}, the right
hand side of Eq.~(\ref{scaling}) must be positive.  We must thus put an upper
bound on $q(\zeta M)^2$: \begin{equation}
q(\zeta M)^2 < {3\pi/2\over \ln(\zeta\xi M\mu)}
\label{limit}
\end{equation}
However, this constraint is in the opposite sense as condition
(\ref{newcondition})  for the removal of the horizon.  In fact, they can
be compatible only if $2M\mu\ln(\zeta\xi M\mu) = 2M\mu\ln(M\mu)
+ 2M\mu\ln(\zeta\xi)< 3\pi/2$.  However, since  $\zeta\xi$ must be a few times
unity, this last inequality can be satisfied only if the Compton
length $1/\mu$ is almost as large as $M$, the black hole's radius.
Nedless to say, the infall of a particle of this size cannot be treated
classically; its evolution in the black hole background is in all cases quantum
mechanical.  Thus we cannot draw the conclusion that the horizon can be removed
by a particle obeying condition (\ref{newcondition}).

Our discussion has been qualitative because the analysis of strongly
coupled $U(1)$ field theory is not yet feasible.  It is clear, however, that
the physics of strongly coupled $U(1)$ is just what is needed to protect
cosmic censorship and the second law.  We find it significant that a classical
black hole requires the help of a quantum effect (vacuum polarization) to
preserve its integrity while absorbing charges.  Perhaps this was to be
expected from the quantum nature of black hole entropy which enters into the
second law of thermodynamics for black holes.

It would obviously be interesting to explore further the question with
lattice simulations of strongly coupled QED to see if  the effective long
range charge is indeed kept small enough to comply with considerations raised
by our discussion.

\vskip 0.3truecm

J. D. B.'s work is supported by a grant from the Israel Science
Foundation which is administered by the Israel Academy of Sciences and
Humanities.  C. R. is supported by the DOE under control number DE-FGO2-85-ER
40231, and also thanks the Racah Institute, Hebrew University for
hospitality.  We both thank Renaud Parentani for a suggestion.

\end{document}